%% file: main.tex
\documentclass[article]{sn-jnl} 

\usepackage[utf8]{inputenc}
\usepackage{graphicx}%
\usepackage{multirow}%
\usepackage{amsmath,amssymb,amsfonts}%
\usepackage{amsthm}%
\usepackage{mathrsfs}%
\usepackage[title]{appendix}%
\usepackage{xcolor}%
\usepackage{textcomp}%
\usepackage{manyfoot}%
\usepackage{booktabs}%
\usepackage{algorithm}%
\usepackage{algorithmicx}%
\usepackage{algpseudocode}%
\usepackage{listings}%

\usepackage{gensymb}
\usepackage[super, sort&compress, comma]{natbib}
\usepackage{float}
\usepackage{fancyhdr}
\usepackage{setspace}
\usepackage{subfiles}
\usepackage{xurl}
\usepackage{bbm}

\usepackage{titlesec}
\titlespacing*{\section}{0pt}{0.5\baselineskip}{0.2\baselineskip}
\titleformat{\subsection}
{\normalfont\large\bfseries}{\thesubsection}{1em}{}
\titlespacing*{\subsection}{0pt}{0.5\baselineskip}{0.2\baselineskip}

\newcommand{\cellfinder}{CellFinder}
\newcommand{\cellsam}{CellSAM}
\newcommand{\cellsamSpecific}{CellSAM-specific}
\newcommand{\cellsamGeneral}{CellSAM-general}

\begin{document}

\setlength{\parindent}{15pt} 

\title{A Foundation Model for Cell Segmentation}



\author[1,3]{\fnm{Uriah} \sur{Israel}}\email{ulisrael@caltech.edu}
\equalcont{These authors contributed equally to this work.}

\author[2,3]{\fnm{Markus} \sur{Marks}}\email{marks@caltech.edu}
\equalcont{These authors contributed equally to this work.}

\author[3]{\fnm{Rohit} \sur{Dilip}}\email{rdilip@caltech.edu}
\equalcont{These authors contributed equally to this work.}

\author[2]{\fnm{Qilin} \sur{Li}}\email{qli2@caltech.edu}

\author[1]{\fnm{Morgan} \sur{Schwartz}}\email{msschwartz@caltech.edu}

\author[1]{\fnm{Elora} \sur{Pradhan}}\email{epradhan@caltech.edu}

\author[1]{\fnm{Edward} \sur{Pao}}\email{epao@caltech.edu}

\author[1]{\fnm{Shenyi} \sur{Li}}\email{sli5@caltech}

\author[1]{\fnm{Alexander} \sur{Pearson-Goulart}}\email{pearsongoulart@gmail.com}

\author[2,3]{\fnm{Pietro} \sur{Perona}}\email{perona@caltech.edu}

\author[3]{\fnm{Georgia} \sur{Gkioxari}}\email{georgia@caltech.edu}

\author[1]{\fnm{Ross} \sur{Barnowski}}\email{rossbar@caltech.edu}

\author[3]{\fnm{Yisong} \sur{Yue}}\email{yyue@caltech.edu}

\author*[1,4]{\fnm{David} \sur{Van Valen}}\email{vanvalen@caltech.edu}

\affil*[1]{\orgdiv{Division of Biology and Biological Engineering}, \orgname{Caltech}}

\affil[2]{\orgdiv{Division of Engineering and Applied Science}, \orgname{Caltech}}

\affil[3]{\orgdiv{Division of Computing and Mathematical Science}, \orgname{Caltech}}

\affil[4]{\orgname{Howard Hughes Medical Institute}}

\abstract{
    Cells are the fundamental unit of biological organization, and identifying them in imaging data - cell segmentation - is a critical task for various cellular imaging experiments. While deep learning methods have led to substantial progress on this problem, models that have seen wide use are specialist models that work well for specific domains. Methods that have learned the general notion of ``what is a cell" and can identify them across different domains of cellular imaging data have proven elusive. In this work, we present CellSAM, a foundation model for cell segmentation that generalizes across diverse cellular imaging data. CellSAM builds on top of the Segment Anything Model (SAM) by developing a prompt engineering approach to mask generation. We train an object detector, CellFinder, to automatically detect cells and prompt SAM to generate segmentations. We show that this approach allows a single model to achieve state-of-the-art performance for segmenting images of mammalian cells (in tissues and cell culture), yeast, and bacteria collected with various imaging modalities. To enable accessibility, we integrate CellSAM into DeepCell Label to further accelerate human-in-the-loop labeling strategies for cellular imaging data. A deployed version of CellSAM is available at \href{https://label-dev.deepcell.org/}{https://label-dev.deepcell.org/}.
    
}

\keywords{cell segmentation, object detection, deep learning, foundation model}

\doublespacing
\maketitle


\section{Introduction}\label{intro}

Accurate cell segmentation is crucial for quantitative analysis and interpretation of various cellular imaging experiments. Modern spatial genomics assays can produce data on the location and abundance of ~$10^1$-$10^2$ protein species and $10^2$-$10^4$ RNA species simultaneously in living and fixed tissues \cite{palla2022spatial, moffitt2022emerging, moses2022museum, hickey2022spatial, ko2022spatiotemporal}. These data shed light on the biology of healthy and diseased tissues but are challenging to interpret. Cell segmentation enables these data to be converted to interpretable tissue maps of protein localization and transcript abundances. Similarly, live-cell imaging provides insight into dynamic phenomena in bacterial and mammalian cell biology. Mechanistic insights into critical phenomena such as the mechanical behavior of the bacterial cell wall \cite{wang2012helical, rojas2018outer}, information transmission in cell signaling pathways \cite{hansen2015limits, hansen2013promoter, tay2010single, regot2014high}, heterogeneity in immune cell behavior during immunotherapy \cite{alieva2023bridging}, and the morphodynamics of development \cite{cao2020establishment} have been gained by analyzing live-cell imaging data. Like their tissue counterparts, cell segmentation is also a key challenge for these experiments, as cells must be segmented and tracked to create temporally consistent records of cell behavior that can be queried at scale.

Significant progress has been made in recent years on the problem of cell segmentation, primarily driven by advances in deep learning \cite{schwartz2023scaling}. Progress in this space has occurred mainly in two distinct but related directions. In the first direction is work that explores the space of deep learning methods that generalize well to cellular imaging data. This includes explorations on deep learning architectures that generalize as well as the representations used to present the notion of what a cell is to a given model \cite{stringer2021cellpose, pachitariu2022cellpose, schmidt2018cell, greenwald2022whole, hollandi2020nucleaizer, graham2019hover, wang2019learn}. The second direction is to work on improving labeling methodology. Cell segmentation is a variant of the instance segmentation problem, which requires pixel-level labels for every object in an image. Creating these labels can be expensive ($10^{-2} - 10^0$ USD/label) \cite{greenwald2022whole, van2016deep}, which provides an incentive to reduce the marginal cost of labeling. A recent improvement to labeling methodology has been human-in-the-loop labeling, where labelers correct model errors rather than produce labels from scratch \cite{greenwald2022whole, pachitariu2022cellpose, Schwartz803205}.

Despite this progress, two critical gaps still need to be addressed. The first is a cell segmentation method that can generalize across diverse cellular images. Existing methods are primarily specialist models - design choices in cellular representation restrict their accuracy to a specific domain. For example, Mesmer's \cite{greenwald2022whole} representation for a cell (cell centroid and boundary) enables good performance in tissue images but would be a poor choice for elongated bacterial cells. Similar trade-offs in representations exist for the current collection of Cellpose models, necessitating the creation of a model zoo \cite{pachitariu2022cellpose}. The second gap is new labeling methodologies that can further reduce the marginal cost of cell labeling. While this cost has been reduced substantially by recent work \cite{greenwald2022whole, Schwartz803205}, reducing this further could increase the amount of labeled imaging data by orders of magnitude.

Recent work in machine learning on foundation models holds promise for providing a complete solution. Foundation models are large deep neural network models (typically transformers \cite{vaswani2017attention}) trained on a large amount of data in a self-supervised fashion with supervised fine-tuning on one or several tasks \cite{bommasani2022opportunities}. Foundation models include the GPT \cite{brown2020language, openai2023gpt4} family of models, which have proven transformative for natural language processing \cite{bommasani2022opportunities} and have been used in other domains, such as biological sequences \cite{lin2023evolutionary}. These successes have inspired similar efforts in computer vision. The Vision Transformer \cite{dosovitskiy2020image} was introduced in 2020 and has since been used as the basis architecture for a collection of vision foundation models \cite{caron2021emerging, oquab2023dinov2, fang2022eva, radford2021learning, alayrac2022flamingo}. One recent foundation model well suited to cellular image analysis needs is the Segment Anything Model (SAM) \cite{kirillov2023segment}. This model uses a Vision Transformer (ViT) to extract information-rich features from raw images. These features are then directed to a module that generates instance masks based on prompts, which can be either spatial (e.g., an object centroid or bounding box) or semantic (e.g., an object’s visual description). Notably, the promptable nature of SAM enabled scalable dataset construction, as preliminary versions of SAM allowed labelers to generate accurate instance masks with 1-2 clicks. The final version of SAM was trained on a dataset of 1 billion masks over 11 million images and demonstrated strong performance on various zero-shot learning tasks. Recent work has attempted to apply SAM to problems in biological and medical imaging, including medical image segmentation \cite{huang2023segment, zhang2023input, lei2023medlsam}, lesion detection in dermatological images \cite{shi2023generalist, hu2023skinsam}, nuclear segmentation in H\&E images \cite{deng2023segment, hörst2023cellvit} and fine-tuned SAM on cellular image data for use in the Napari software package \cite{Archit2023.08.21.554208}.

While promising, these studies reported challenges adapting SAM to these new use cases \cite{huang2023segment, Archit2023.08.21.554208}. These challenges include reduced performance and uncertain boundaries when transitioning from natural to medical images. Cellular images contain additional complications -- they can involve different imaging modalities (e.g., phase microscopy vs. fluorescence microscopy), thousands of objects in a field of view (as opposed to dozens in a natural image), uncertain and noisy boundaries (artifacts of projecting 3D objects into a 2D plane) \cite{Archit2023.08.21.554208}. In addition to these challenges, SAM's default prompting strategy does not allow for accurate inference for cellular images. Currently, the automated prompting of SAM uses a uniform grid of points to generate masks, an approach poorly suited to cellular images given the wide variation of cell densities. More precise prompting (e.g., a bounding box or mask) requires prior knowledge of cell locations. This creates a weak tautology - SAM can find the cells provided it knows a priori where they are. This limitation makes it challenging for SAM to serve as a foundation model for cell segmentation - it can accelerate labeling but still requires human input for inference. A solution to this problem would enable SAM-like models to serve as foundation models and knowledge engines, as they could accelerate the generation of labeled data, learn from them, and make that knowledge accessible to life scientists via inference. 
 
\begin{figure*}
        \centering
        \includegraphics[width=\linewidth]{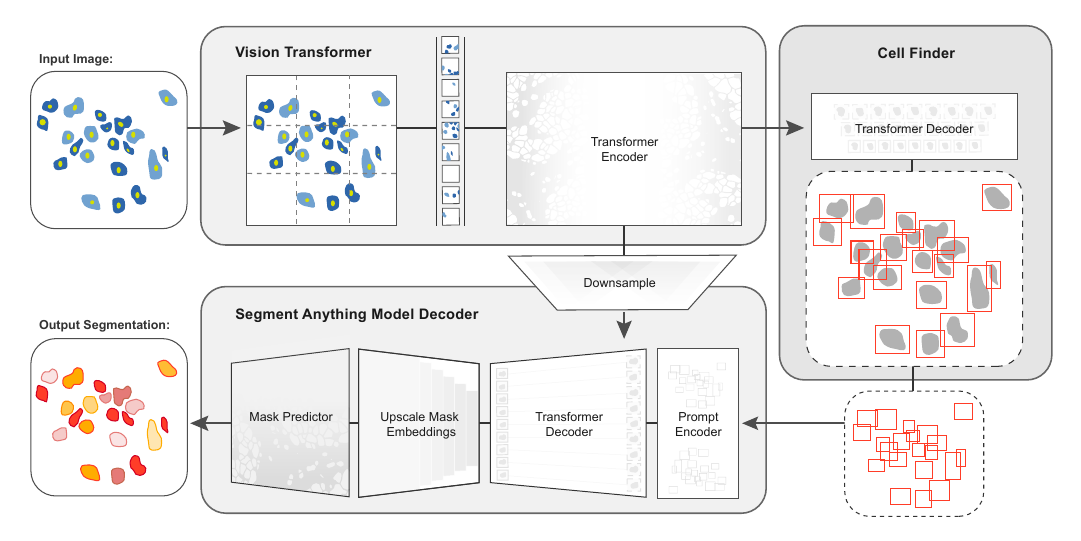}
        \vspace{-0.23in}
        \caption{\textbf{\cellsam: a foundational model for cell segmentation.}
        \cellsam combines SAM's mask generation and labeling capabilities with an object detection model to achieve automated inference. Input images are divided into regularly sampled patches and passed through a transformer encoder (e.g., a ViT) to generate information-rich image features. These image features are then sent to two downstream modules. The first module, \cellfinder, decodes these features into bounding boxes using a transformer-based encoder-decoder pair. The second module combines these image features with prompts to generate masks using SAM's mask decoder. \cellsam~integrates these two modules using the bounding boxes generated by \cellfinder~as prompts for SAM. \cellsam~is trained in two stages, using the pre-trained SAM model weights as a starting point. In the first stage, we train the ViT and the \cellfinder~model together on the object detection task. This yields an accurate \cellfinder~but results in a distribution shift between the ViT and SAM's mask decoder. The second stage closes this gap by fixing the ViT and SAM mask decoder weights and fine-tuning the remainder of the SAM model (i.e., the model neck) using ground truth bounding boxes and segmentation labels.}
        
        \label{fig:main_overview}
    \end{figure*}

In this work, we developed \cellsam, a foundation model for cell segmentation (Fig. \ref{fig:main_overview}). \cellsam~extends the SAM methodology to perform automated cellular instance segmentation. To achieve this, we first assembled a comprehensive dataset for cell segmentation spanning five different morphological archetypes. To automate inference with SAM, we took a prompt engineering approach and explored the best ways to prompt SAM to generate high-quality masks. We observed that bounding boxes consistently generated high-quality masks compared to alternative approaches. We further identified a compute-efficient method to fine-tune SAM to achieve even better performance. To facilitate automated inference through prompting, we developed \cellfinder, a transformer-based object detector that uses the Anchor DETR framework. Within \cellsam, \cellfinder~and SAM shares the same ViT backbone; the bounding boxes generated by \cellfinder~are then used as prompts for SAM, enumerating masks for all the cells in an image. We trained \cellsam~on a large, diverse corpus of cellular imaging data, enabling it to achieve state-of-the-art (SOTA) performance on nine datasets. We also evaluated \cellsam's zero-shot performance using a held-out dataset \cite{edlund2021livecell}, demonstrating that it outperforms existing methods for zero-shot segmentation. The datasets described in this work are available at \url{https://deepcell.readthedocs.io/en/master/data-gallery/}; a deployed version of \cellsam~is available at our lab's web portal \href{https://deepcell.org}{https://deepcell.org}.
    
\section{Results}\label{results}
\subsection{Construction of a dataset for general cell segmentation}\label{dataset-results}
A significant challenge with existing cellular segmentation methods is their inability to generalize across various imaging modalities and cell morphologies. To address this, we curated a dataset from the literature containing 2D images of various cell morphologies (mammalian cells in tissues and adherent cell culture, yeast cells, bacterial cells, and mammalian cell nuclei) and imaging modalities (fluorescence, brightfield, phase contrast, hematoxylin \& eosin staining, and mass cytometry imaging). For each ingested dataset, we inspected them for data leaks between training and testing splits and removed them when present. Our final dataset consisted of TissueNet~\cite{greenwald2022whole}, DeepBacs~\cite{spahn2022deepbacs}, BriFiSeg~\cite{mathieu2022brifiseg}, Cellpose~\cite{stringer2021cellpose, pachitariu2022cellpose}, Omnipose~\cite{cutler2021omnipose, cutler2022omnipose}, YeastNet~\cite{kim2014yeastnet}, YeaZ~\cite{dietler2020yeaz}, the 2018 Kaggle Data Science Bowl dataset (DSB)~\cite{caicedo2019nucleus}, and an internally collected dataset of phase microscopy images across eight mammalian cell lines (Phase400). For evaluation, we group these datasets into four types: Tissue, Cell Culture, Bacteria, and Yeast. As the DSB~\cite{caicedo2019nucleus} comprises cell nuclei that span several of these types, we evaluate it separately and refer to it as Nuclear. While our method focuses on whole-cell segmentation, we included DSB~\cite{caicedo2019nucleus} because cell nuclei are often used as a surrogate when the information necessary for whole-cell segmentation (e.g., cell membrane markers) is absent from an image. A summary of the dataset is shown in Figure \ref{fig:main_results}a. To evaluate \cellsam's zero-shot performance, we used a held-out LIVECell \cite{edlund2021livecell} dataset. A detailed description of data sources and pre-processing steps can be found in the Appendix~\ref{sec:appendix:data}.

\subsection{Bounding boxes are accurate prompts for cell segmentation with SAM}\label{prompting-results}
\begin{figure*}[h]
        \centering
        \includegraphics[width=\linewidth]{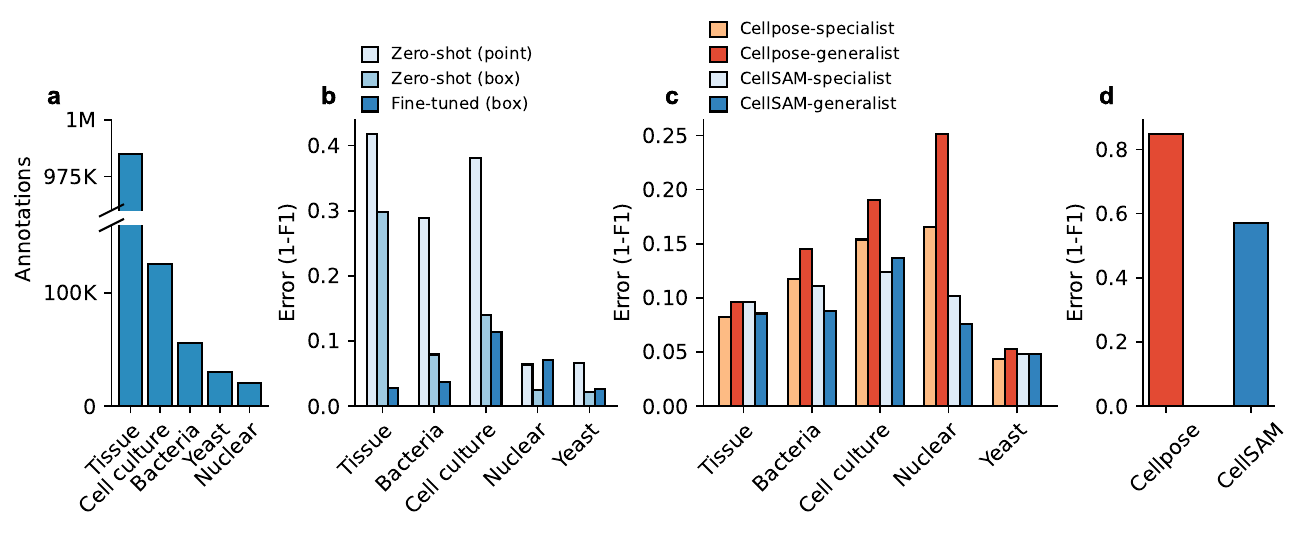}
        \includegraphics[width=\linewidth]{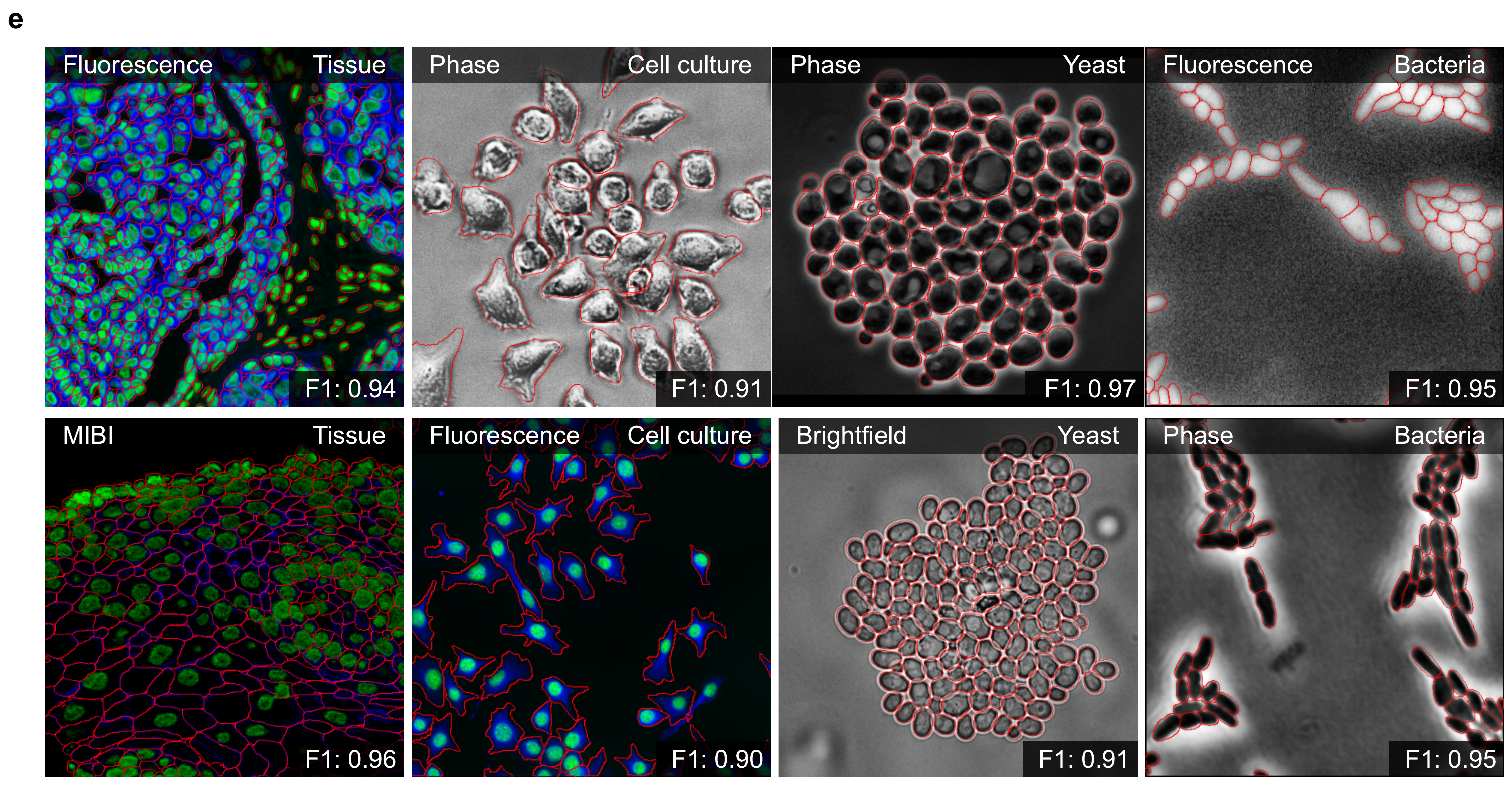}
        
        \caption{\textbf{\cellsam~is a strong generalist model for cell segmentation.} a) For training and evaluating \cellsam, we curated a diverse cell segmentation dataset from the literature. The number of annotated cells is given for each data type. Nuclear refers to a heterogeneous dataset (DSB) \cite{caicedo2019nucleus} containing nuclear segmentation labels. b) Zero-shot (ZS) and fine-tuned mask generation error (1- F1 score) for SAM when using point and bounding box prompts. All prompting in this figure was done with ground truth prompts. The best performance is achieved with bounding box prompts and fine-tuning. c) Segmentation performance for \cellsam~and Cellpose on different data types. We compare the segmentation error (1-F1) for models that were trained as specialists (e.g., on one dataset) or generalists (the full dataset). Models were trained for a similar number of steps across all datasets. We observed that \cellsamGeneral has a lower error than Cellpose-general on almost all tested datasets. Further, we observed that generalist training improved \cellsam's performance over specialist training; the reverse was true for Cellpose. d) Zero-shot performance of \cellsamGeneral~and Cellpose-General on the LIVECell dataset. Here, we show greater than 4x segmentation performance on an unseen dataset.  e) Qualitative results of \cellsam~segmentations for different data and imaging modalities. Predicted segmentations are outlined in red.
        }
        \label{fig:main_results}
    \end{figure*}

For accurate inference, SAM needs to be provided with approximate information about the location of cells in the form of prompts. To better engineer prompts, we first assessed SAM's ability to generate masks when provided prompts derived from ground truth labels - either point prompts (derived from the cell's center of mass) or bounding box prompts. For these tests, we used the pre-trained model weights that were publicly released\cite{kirillov2023segment}. Our benchmarking results are shown in Figure \ref{fig:main_results}b and revealed that bounding boxes had significantly higher zero-shot performance than point prompting, although both approaches struggled with Tissue imaging data. To improve SAM's mask generation ability for cellular image data, we explored fine-tuning SAM on our compiled data to help it bridge the gap from natural to cellular images. During these fine-tuning experiments, we observed that fine-tuning all of SAM was unnecessary; instead, we only needed to fine-tune the layers connecting SAM's ViT to its decoder, the model neck, to achieve good performance. All other layers can be frozen. Fine-tuning SAM in this fashion led to a model capable of generating high-quality cell masks when prompted by ground truth bounding boxes, as seen in Figure \ref{fig:main_results}b.

\subsection{\cellfinder~and \cellsam~enable accurate and automated cell segmentation}\label{cellfinder-results}
Given that bounding box prompts yield accurate segmentation masks from SAM across various datasets, we sought to develop an object detector that could generate prompts for SAM in an automated fashion. Given that our zero-shot experiments demonstrated that ViT features can form robust internal representations of cellular images, we reasoned we could build an object detector on top of the image features generated by SAM's ViT. Previous work has explored this space and demonstrated that ViT backbones can achieve SOTA performance on natural images \cite{li2022exploring, lin2014microsoft}. For our object detection module, we use the Anchor DETR framework \cite{wang2022anchor}, using the same ViT backbone as the SAM module; we call this object detection module \cellfinder. Anchor DETR is well suited for object detection in cellular images because it formulates object detection as a set prediction task. This allows it to - in theory - perform cell segmentation in images that are densely packed or contain overlapping objects, common occurrences in cellular imaging data. These failure modes are challenging to address with existing methods. Bounding box methods (e.g., the R-CNN family \cite{girshick2014rich, ren2016faster}) rely on non-maximum suppression, leading to poor performance in this regime. Methods that frame cell segmentation as a dense, pixel-wise prediction task (e.g., Mesmer\cite{greenwald2022whole} and Cellpose\cite{stringer2021cellpose}) assume that each pixel can be uniquely assigned to a single cell and cannot handle overlapping objects.

We train \cellsam~in two stages; the full details can be found in Appendices \ref{appendix:cellsam}. In the first stage, we train \cellfinder~on the object detection task. We convert the ground truth cell masks into bounding boxes and train the ViT backbone and the \cellfinder~module. Once \cellfinder~is trained, we freeze the model weights of the ViT and fine-tune the SAM module as described above. This accounts for the distribution shifts in the ViT features that occur during the \cellfinder~training. Once training is complete, we use \cellfinder~to prompt SAM's mask decoder. We refer to the collective method as \cellsam; Figure \ref{fig:main_overview} outlines an image's full path through \cellsam~during inference. We benchmark \cellsam's performance using a suite of metrics (Figure \ref{fig:main_results}c and \ref{fig:main_results}d and Supplemental Figure \ref{supp:fig:data_specific_cellsam_metrics}) and find that it outperforms Cellpose models trained on comparable datasets. We highlight two features of our benchmarking analyses below.
\begin{itemize}
    \item \textbf{\cellsam~is a strong generalist model}. Generalization across cell morphologies and imaging datasets has been a significant challenge for deep learning-based cell segmentation algorithms. To evaluate \cellsam's generalization capabilities, we compared its performance to \cellsam~and Cellpose models trained as specialists (e.g., on a single dataset) or generalists (e.g., on the entire dataset). Consistent with the literature, we observed that Cellpose's performance degraded when trained as a generalist (Figure \ref{fig:main_results}c), as specialist Cellpose models had a higher F1 score across all datasets. We observed that the reverse was true for \cellsam; the F1 score remained the same or improved in four of the five data categories and across seven of the nine datasets (Figure \ref{fig:main_results} and Supplemental Figure \ref{supp:fig:data_specific_cellsam_metrics}).
    \item \textbf{\cellsam~achieves SOTA zero-shot performance}. To further evaluate \cellsam's capacity for generalization, we evaluated its performance on an entirely unseen dataset, LIVECell \cite{edlund2021livecell}, without further fine-tuning. When compared against the Cellpose-generalist model, we find that \cellsam's zero-shot segmentation performance is considerably better, albeit still not accurate enough to be used in real-world settings. We note that some of the poor reported performance is due to label errors in the LIVECell dataset\cite{pachitariu2022cellpose}.
\end{itemize}



\section{Discussion}
Cell segmentation is a critical task for cellular imaging experiments. While deep learning methods have made substantial progress in recent years, there remains a need for methods that can generalize across diverse images and further reduce the marginal cost of image labeling. In this work, we sought to meet these needs by developing \cellsam, a foundation model for cell segmentation. Transformer-based methods for cell segmentation are showing promising performance. \cellsam~builds on these works by integrating the mask generation capabilities of SAM with transformer-based object detection to empower both scalable image labeling and automated inference. We trained \cellsam~on a diverse dataset curated from the literature. Our benchmarking demonstrated that \cellsam~achieves SOTA performance on cell segmentation and that this performance is aided by our attempts to create a general segmentation model. Given its utility in image labeling and accuracy during inference, we believe \cellsam~is a valuable contribution to the field and will help create the data infrastructure required for cellular imaging's AI-powered future.

The work described here has importance beyond aiding life scientists with cell segmentation. First, foundation models are immensely useful for natural language and vision tasks and hold similar promise for the life sciences - provided they are suitably adapted to this new domain. We can see several uses for \cellsam~that might be within reach of future work. First, given its generalization capabilities, it is likely that \cellsam~has learned a general representation for the notion of ``cells" used to query imaging data. These representations might serve as an interface between imaging data and other modalities (e.g., single-cell RNA Sequencing), provided there is suitable alignment between cellular representations for each domain \cite{ zhang2022graph, yang2021multi}. Second, much like what has occurred with natural images, we foresee that the integration of natural language labels in addition to cell-level labels might lead to vision-language models capable of generating human-like descriptors of cellular images with entity-level resolution\cite{radford2021learning}. Third, the generalization capabilities may enable the standardization of cellular image analysis pipelines across all the life sciences. If the accuracy is sufficient, microbiologists and tissue biologists could use the same collection of foundation models for interpreting their imaging data even for challenging experiments \cite{shah2017seqfish, dar2021spatial}. Last, new efforts seek to generate AI scientists capable of generating hypotheses and exploring them through the design and execution of new experiments \cite{jablonka202314}. Foundation models like \cellsam~could contribute to this vision by serving as this scientist's ``eyes", converting complex imaging data to structured knowledge that can be operationalized.

While the work presented here highlights the potential foundation models hold for cellular image analysis, much work remains to be done for this future to manifest. Extension of this methodology to 3D imaging data is essential; recent work on memory-efficient attention kernels \cite{nguyen2023hyenadna} will aid these efforts. Exploring how to enable foundation models to leverage the full information content of images (e.g., multiple stains, temporal information for movies, etc.) is an essential avenue of future work. Expanding the space of labeled data remains a priority - this includes images of perturbed cells and cells with more challenging morphologies (e.g., neurons). Data generated by pooled optical screens \cite{feldman2019optical} may synergize well with the data needs of foundation models. Compute-efficient fine-tuning strategies must be developed to enable flexible adaptation to new image domains. Lastly, prompt engineering is a critical area of future work, as it is critical to maximizing model performance. The work we presented here can be thought of as prompt engineering, as we leverage \cellfinder~to produce bounding box prompts for SAM. As more challenging labeled datasets are incorporated, the nature of the ``best" prompts will likely evolve. Finding the best prompts for these new data, rather than the best vision pipelines, is a task that will likely fall on both the computer vision and life science communities. 

\clearpage

{\small
\bibliographystyle{IEEEtran}
\bibliography{main}
}

\clearpage
\section*{Declarations}

\subsection*{Code Availability}
The Van Valen lab is a strong believer in open-source software development. A github repository will be made available at \url{https//github.com/vanvalenlab} as soon as a library with satisfactory software engineering standards is compiled.

\subsection*{Data Availability}
The dataset used to develop \cellsam~is available at \url{https://deepcell.readthedocs.io/en/master/data-gallery/index.html} for non-profit use.

\subsection*{Author Contributions}
UI, MM, YY, and DVV conceived the project; UI, MM, QL, YY, and DVV performed algorithm design for \cellfinder~and \cellsam; MM implemented the \cellsam~ architecture; UI, MM and QL implemented \cellfinder. UI and MM carried out the experiments and evaluations of the method. GG and PP provided input for developing \cellfinder; QL and UI performed model benchmarking; QL and RD developed data pipelines, RD developed the computational infrastructure for model training; RD, EP, EP, MS, QL, and RB performed data engineering; SL, APG, RD, and RB performed the DeepCell Label-\cellsam~integration, RB and DVV supervised the software engineering, DVV supervised the project.

\subsection*{Acknowledgements}
We thank Leeat Keren, Noah Greenwald, Sam Cooper, Jan Funke, Uri Manor, Joe Horsman, Michael Baym, Paul Blainey, Ian Cheeseman, Manuel Leonetti, Changhua Yu, Neehar Kondapaneni, and Elijah Cole for valuable conversations and insightful feedback. We also thank William Graf, Geneva Miller, and Kevin Yu, whose time in the Van Valen lab established the infrastructure and software tools that made this work possible. We thank Nader Khalil, Alec Fong, and the entire Brev.dev team for their support in establishing the computational infrastructure required for this work. We utilized images of the HeLa cell line in this research. Henrietta Lacks and the HeLa cell line established from her tumor cells without her knowledge or consent in 1951 has significantly contributed to scientific progress and advances in human health. We are grateful to Lacks, now deceased, and the Lacks family for their contributions to biomedical research. This work was supported by awards from the Shurl and Kay Curci Foundation (to DVV), the Rita Allen Foundation (to DVV), the Susan E. Riley Foundation (to DVV), the Pew-Stewart Cancer Scholars program (to DVV), the Gordon and Betty Moore Foundation (to DVV), the Schmidt Academy for Software Engineering (to SL), the Michael J. Fox Foundation through the Aligning Science Across Parkinson's consortium (to DVV), the Heritage Medical Research Institute (to DVV), the National Institutes of Health New Innovator program (DP2-GM149556) (to DVV), the National Institutes of Health HuBMAP consortium (OT2-OD033756) (to DVV), and the Howard Hughes Medical Institute Freeman Hrabowski Scholars program (to DVV). National Institutes of Health (R01-MH123612A) (to PP). NIH/Ohio State University (R01-DC014498) (to PP). Chen Institute (to PP). The Emerald Foundation and Black in Cancer (to UI). Caltech Presidential Postdoctoral Fellowship Program (PPFP) (to UI).

\subsection*{Disclosures}
 David Van Valen is a co-founder and Chief Scientist of Barrier Biosciences and holds equity in the company. All other authors declare no competing interests.
 
\clearpage

\input{supplement}

\end{document}

%% file: supplement.tex
\appendix

\section{Dataset Construction}
    \label{sec:appendix:data}

    To train CellSAM, we combined nine separate datasets spanning a variety of modalities: TissueNet~\cite{greenwald2022whole}, DeepBacs~\cite{spahn2022deepbacs}, BriFiSeg~\cite{mathieu2022brifiseg}, Cellpose~\cite{stringer2021cellpose, pachitariu2022cellpose}, Omnipose~\cite{cutler2021omnipose, cutler2022omnipose}, YeastNet~\cite{kim2014yeastnet}, YeaZ~\cite{dietler2020yeaz}, the 2018 Kaggle Data Science Bowl (DSB)~\cite{caicedo2019nucleus}, and an internally collected dataset of phase microscopy images across eight mammalian cell lines (Phase400). The LIVECell \cite{edlund2021livecell} dataset was held out for zero-shot testing. Our collective dataset included images across multiple imaging modalities (brightfield, phase contrast, h\&e staining, fluorescence, and mass cytometry), imaging targets (histology sections, yeast, cell culture, bacteria, nuclei), length scales, and morphologies. During preprocessing, every image in our dataset was normalized using Contrast Limited Adaptive Histogram Equalization (CLAHE)\cite{pizer1987adaptive} with a kernel size of 128 pixels. We treated nuclear and whole-cell channels as green and blue channels in an RGB image, respectively, and the red channel is always blank. We moved the green channel to blue for nuclear-only datasets (i.e., BriFiSeg and DSB) to keep the blue channel always non-empty.
    
    If available, we used pre-determined train/val/test splits for each dataset; otherwise, we introduced 80-10-10\% data splits. For datasets with multiple fields of view of the same object set, we required all FOVs to belong to the same split. We defer all duplicated samples to the train split for published datasets with a pre-existing data leak. Our assembled dataset uses a fixed image size of 512 by 512 pixels. Images shorter than 512 pixels on either axis are zero-padded up to 512. For images with more than 512 pixels on either axis, we tiled them to 512 by 512 pixels with a 25\% overlap and filled the empty regions with zeros. Any cropped images without valid annotations were removed. We follow a widely used annotation scheme for labeling our masks, with zero representing the background and unique positive integers representing different objects. While this format precludes accurate segmentation of overlapping objects, labels of this kind were not present in the dataset we compiled. We filtered out invalid cell labels if 1) the label contained disjoint regions, typically caused by random mouse clicking; 2) the label has only a 1-pixel height or width. The cropped images with filtered annotations are used for training, validation, and testing. LIVECell\cite{edlund2021livecell} annotations were converted from the COCO format to this labeling format for consistency. We used cellpose's\cite{pachitariu2022cellpose} pre-processing function \texttt{livecell\_ann\_to\_masks()} to remove overlapping regions. To match the phase contrast cell size in the training set, we rescale the LIVECell images by 2.0 before the standard image preprocessing pipeline. We use scikit-image\cite{scikit-image} \texttt{skimage.transform.rescale()} function with \texttt{bicubic} interpolation for images and \texttt{nearest} interpolation for annotation masks.
    
    To summarize our dataset format, our images have RGB channels with a fixed size of 512 by 512 pixels, stored in shape (3, 512, 512) float32 array in the range [0, 1]. The blue channel was the main channel, reserved for whole-cell images, and always non-empty. The green channel was the supplementary channel used for nuclear images but could be empty. The blue channel was used if only nuclear images were available (e.g., for the DSB dataset). The red channel was always empty. Our label masks had the same height and width as the images, stored with the shape (1, 512, 512) int32 in the range [0, number of objects]. We stored the processed dataset in two formats. The numpy npy format was used for \cellsam~fine-tuning and model evaluation. The COCO format \cite{lin2014microsoft} was used for \cellfinder~and ViT backbone training.

\section{CellSAM Architecture.}\label{appendix:cellsam}
    
    We adapted Anchor DETR \cite{wang2022anchor} for the object detector for \cellsam~(\cellfinder). This choice was motivated by Anchor DETR being non-maximum suppression (NMS)\cite{hosang2017learning} free. NMS suppresses bounding boxes with a high amount of overlap to remove duplicate detections. While this works well for natural images, cellular images often have tightly clustered objects, and NMS-based methods such as the R-CNN family \cite{ren2016faster, girshick2014rich} can suffer from a low recall in this setting. We replaced the Anchor DETR's ResNet~\cite{he2016deep} backbone with the vision transformer (ViT)~\cite{dosovitskiy2020image} from the SAM model \cite{kirillov2023segment}; specifically, we used the base-sized ViT (ViT-B).
    
    As the maximum number of cells per image is generally no more than 1000, we increased the number of queries $q$ to 3500, 3.5 times the maximum number of cells, based on Fig. 12 in DETR\cite{carion2020end}, which provides an estimate of the number of queries needed for a DETR method to detect all objects. We used only one pattern $p$ for the Anchor generation as most objects in cellular detection are usually of similar scale. 
    
    \textit{Training \cellfinder} We used a base learning rate of $10^{-4}$ for the Anchor DETR head and $10^{-5}$ for the SAM-ViT backbone. We use weight decay of $10^{-4}$ and clip norm of $0.1$. We apply a dropout of $0.1$. We use AdamW\cite{loshchilov2017decoupled} with a step-wise learning rate scheduler that drops the learning rate by $10\%$ after $70\%$ of the epochs. We train \cellfinder~for 500 epochs (1000 for smaller datasets) with a batch size of $2$ across 16 GPUs.
    
    \textit{Finetuning \cellsam} After we trained \cellfinder~with the SAM-ViT backbone, the SAM-ViT output features were no longer aligned with the rest of the model (i.e., the prompt encoder and mask decoder). To close this distribution gap, we froze the SAM-ViT (such that it continues to function well with \cellfinder) and trained the neck of the SAM model. The neck is a 2D-convolutional neural network that embeds the ViT features (e.g., $768$ for SAM-ViT-B) to a $256$-dimensional embedding that is then used as the primary feature vector for the rest of the model (prompt embedding and mask decoder). We trained this neck using ground-truth bounding boxes as inputs and segmentation masks of individual cells as targets. We used a learning rate of $10^{-4}$ and weight decay of $10^{-4}$ for this training. We also used AdamW\cite{loshchilov2017decoupled} for this training and did not clip the gradient. 

\subsection{Inference}

    At inference, we followed the following workflow. First, the input was passed through the Anchor DETR fine-tuned ViT-B. This resulted in an embedding dimension of $768$. This embedding was then passed as an input to two parts of \cellsam, 1) the trained Anchor DETR module (\cellfinder) and 2) the fine-tuned neck, which is a 2D convolutional network reducing the embedding dimensionality further to $256$. The bounding box outputs of \cellfinder~were then sent into the prompt encoder, resulting in the prompt embedding. The prompt embeddings and neck embedding were then passed to the mask decoder, which outputs pixel-wise probabilities for the cell and another IoU-based confidence value for the prediction as a whole. This results in a tensor of shape $N \times W \times H$, where $N$ corresponds to the number of cells predicted. This tensor was processed with a sigmoid and a threshold operation, resulting in binarized images. Depending on the metric used, we either use this tensor directly together with the $N$ scores (specifically for computation of the coco AP @ 0.5 IoU) or we compute the argmax over the cell dimension $N$ to generate a tensor $W \times H$, where each pixel corresponds to an integer that is unique for each cell.
    
    \textit{Thresholding.} Given \cellsam's model architecture, we have three different thresholds at inference time. First, we had a threshold on the bounding boxes generated by \cellfinder, which we set to $0.4$ across all datasets. After the boxes were passed through the Mask Decoder, we had an overall mask score outputted by the IoU prediction head of the Mask Decoder, which we set to $0.5$. Lastly, we thresholded the mask decoder output after applying the sigmoid function to each pixel, which we set at $0.5$.

    \textit{\cellsam~Postprocessing.} We use the same postprocessing steps that are used by SAM\cite{kirillov2023segment}. This consisted of hole filling and island removal for each predicted cell.
    
\subsubsection{Model Implementation and Training}

    \cellsam~is implemented in pytorch\cite{NEURIPS20199015}. For \cellfinder~we modify the official Anchor DETR repo\footnote{\href{https://github.com/megvii-research/AnchorDETR}{https://github.com/megvii-research/AnchorDETR}}. For \cellsam, we modify the official Segment Anything repo\footnote{\href{https://github.com/facebookresearch/segment-anything}{https://github.com/facebookresearch/segment-anything}}. We use pytorch lightning \cite{FalconPyTorchLightning2019} to scale the training. Prototyping was done using NVIDIA's RTX 4090. We used machines with either NVIDIA A6000s or A100s (40GB and 80GB versions) for the experiments in the paper.

\section{Benchmarking}

    We benchmarked the performance of \cellsam~ models against Cellpose\cite{stringer2021cellpose, pachitariu2022cellpose} trained on our compiled datasets.
    
    \subsection{Cellpose Model Training.}  We follow the hyper-parameters described in the original paper\cite{pachitariu2022cellpose} to train specialist and generalist Cellpose models from scratch. We use the SGD optimizer with a weight decay of $10^{-4}$ and a batch size of eight. We train each model for 300 epochs with a base learning rate of $0.1$. The learning rate increases linearly from 0 to 0.1 over the first ten epochs, then decreases by a factor of two every five epochs after the 250th epoch. The main channel (\texttt{--chan}) is 3 (blue), and the supplementary channel (\texttt{--chan2}) is 2 (green). Other hyper-parameters were kept at the default setting. We trained each model on a single NVIDIA A6000 GPU with 11.4GB GPU Memory utilization. In total, we train nine specialist models and one generalist model.

    \subsection{Metrics}
    We used the Metrics package present in the DeepCell library\cite{greenwald2022whole, Schwartz803205}, which is a set of tools for object-level evaluation of cell segmentations. Predictions that match the ground truth labels (determined by a mask IoU $\ge$ 0.6) are true positives (TP), predictions with no matching ground truth labels are false positives (FP), and ground truth labels without a valid match are false negatives. We compute the recall, precision, and F1 scores using the following formulas:
    \begin{itemize}
        \item Recall: ${\mathrm{recall}}=\frac{{\mathrm{TP}}}{\mathrm{{TP+FN}}}$.
        \item Precision: ${\mathrm{precision}}=\frac{{\mathrm{TP}}}{\mathrm{{TP+FP}}}$.
        \item F1: $F_{1}=\frac{2 \times \mathrm{precision} \times \mathrm{recall}}{\mathrm{precision} + \mathrm{recall}}$.
    \end{itemize}
    Details of the implementation of these metrics are described in prior work\cite{Schwartz803205}.

    We also used the COCO evaluation metrics \cite{lin2014microsoft} during \cellfinder's development. The COCO metrics are a widely used benchmark for assessing the object-level quality of object detection and instance segmentation methods. These metrics report Average Precision (AP), the area under the Precision-Recall curve for a given object class. In our case, we only had a single object class - cells. The AP is computed for different IoU thresholds, ranging from 0.5 to 0.95, with a step size of 0.05. We report the mean AP across all IoU thresholds, denoted as \texttt{mAP}, as well as the AP at IoU=0.5, denoted as \texttt{AP50}, to quantify \cellfinder's performance. Because the object density is much higher in cellular images than in natural images, we modified the limit for the maximum number of detections from 100 to 10,000. We also fed the actual confidence score per binary prediction of the \cellsam~model to the COCO evaluator. For the Cellpose models, we used a fixed confidence score of 1.0.

\clearpage

\renewcommand{\thefigure}{S\arabic{figure}}
\setcounter{figure}{0}
\begin{figure*}
    \centering
    \includegraphics[width=\linewidth]{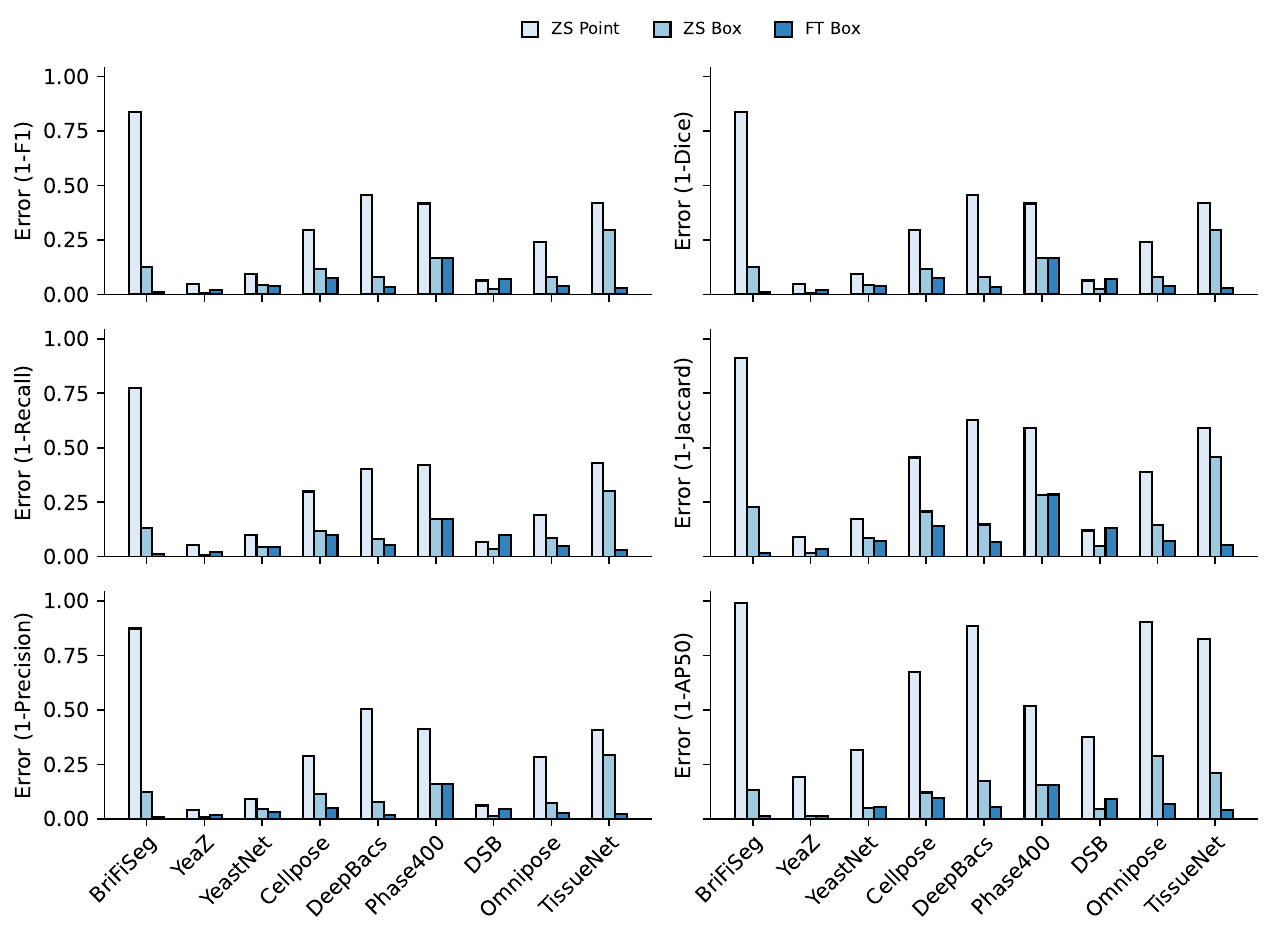}
    \vspace{-0.23in}
    \caption{\textbf{Per dataset performance comparing zero-shot point prompting, zero-shot box prompting, and fine-tuned box prompting} across a suite of metrics from the DeepCell package, and additionally, we included the AP50 from the COCO metrics. We show the error rate (1-metric) on these bar plots. We demonstrate \cellsamSpecific and \cellsamGeneral superior performance across multiple datasets and multiple evaluation metrics. }
    \label{supp:fig:data_zs_ft_metrics}
\end{figure*}

\begin{figure*}
    \centering
    \includegraphics[width=\linewidth]{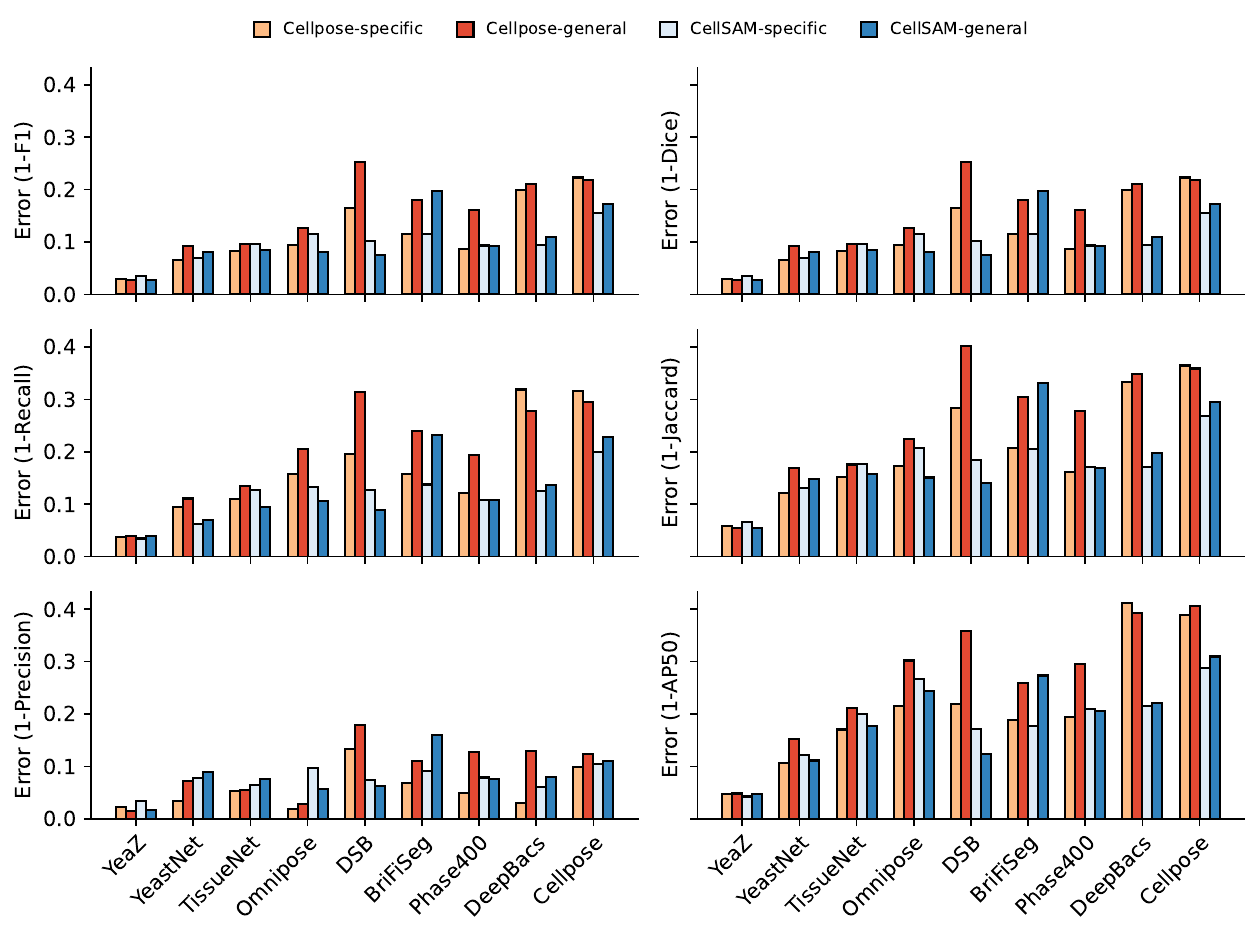}
    \vspace{-0.23in}
    \caption{\textbf{Per dataset performance} across a suite of metrics from the DeepCell package, and additionally, we included the AP50 from the COCO metrics. We show the error rate (1-metric) on these bar plots. We demonstrate \cellsamSpecific and \cellsamGeneral superior performance across multiple datasets and evaluation metrics. }
    \label{supp:fig:data_specific_cellsam_metrics}
\end{figure*}



